\documentclass[a4paper,reqno]{amsart}
\usepackage{color}
\usepackage{enumerate}
\usepackage{float}
\usepackage{amsmath, amsthm, amssymb, comment, array,accents}
\usepackage{pgfplots}
\usepackage[latin1]{inputenc}
\usepackage{pdfsync}
\usepackage{booktabs,bm}
\usepackage{url,fancyref,hyperref}
\hypersetup{colorlinks,citecolor=red,urlcolor=blue,linkcolor=black,filecolor=black}
\usepackage[toc]{appendix}
\usepackage{marginnote}

\usepackage{longtable}
\newcolumntype{L}{>{$}l<{$}}
\newtheoremstyle{mytheoremstyle} 
    {\topsep}                    
    {\topsep}                    
    {\itshape}                   
    {}                           
    {\bfseries}                   
    {:}                          
    {.5em}                       
    {}  

\newtheoremstyle{mydefinitionstyle} 
    {\partopsep}                    
    {\partopsep}                    
    {\normalfont}                   
    {}                           
    {\bfseries}                   
    {:}                          
    {.5em}                       
    {}  

\makeatletter
\def\th@plain{%
  \thm@notefont{}
  \itshape 
}
\def\th@mydefinitionstyle{%
  \thm@notefont{}
  \normalfont 
}
\makeatother
\let\oldsqrt\sqrt
\def\sqrt{\mathpalette\DHLhksqrt}
\def\DHLhksqrt#1#2{%
\setbox0=\hbox{$#1\oldsqrt{#2\,}$}\dimen0=\ht0
\advance\dimen0-0.2\ht0
\setbox2=\hbox{\vrule height\ht0 depth -\dimen0}%
{\box0\lower0.4pt\box2}}

\theoremstyle{mytheoremstyle}
\newtheorem{theorem}{Theorem}[section] 
\newtheorem{prop}[theorem]{Proposition} 
\theoremstyle{mydefinitionstyle}
\newtheorem{defn}[theorem]{Definition}

\newtheorem{remark}[theorem]{Remark} 
\newtheorem{lemma}[theorem]{Lemma} 
\newtheorem{cor}[theorem]{Corollary}

\newcommand{\be}{\begin{equation}}
\newcommand{\ee}{\end{equation}}
\newcommand{\bal}{\begin{align*}}
\newcommand{\eal}{\end{align*}}
\newcommand{\ba}{\begin{array}}
\newcommand{\ea}{\end{array}}
\newcommand{\bmx}{\begin{bmatrix}}
\newcommand{\emx}{\end{bmatrix}}
\newcommand{\ben}{\begin{equation*}}
\newcommand{\een}{\end{equation*}}
\newcommand{\beq}{\begin{eqnarray}}
\newcommand{\eeq}{\end{eqnarray}}
\newcommand{\beqn}{\begin{eqnarray*}}
\newcommand{\eeqn}{\end{eqnarray*}}
\newcommand{\bseq}{\begin{subequations}}
\newcommand{\eseq}{\end{subequations}}
\newcommand{\bmat}{\begin{pmatrix}}
\newcommand{\emat}{\end{pmatrix}}
\newcommand{\bem}{\begin{enumerate}}
\newcommand{\eem}{\end{enumerate}}
\newcommand{\bd}{\begin{description}}
\newcommand{\ed}{\end{description}}
\newcommand{\bl}{\begin{itemize}}
\newcommand{\el}{\end{itemize}}
\newcommand{\bc}{\begin{cases}}
\newcommand{\ec}{\end{cases}}

\newcommand{\sn}{\operatorname{sn}}

\newcommand{\al}{\alpha}
\newcommand{\gm}{\gamma}
\newcommand{\de}{\delta}
\newcommand{\lam}{\lambda}
\newcommand{\ep}{\epsilon}

\newcommand{\sig}{\sigma}

\newcommand{\om}{\omega}

\newcommand{\tht}{\theta}

\newcommand{\De}{\Delta}

\newcommand{\Om}{\Omega}

\newcommand{\Fb}{\overline{F}}

\newcommand{\sbar}{\overline{s}}
\newcommand{\su}{\underline{s}}

\newcommand{\ub}{\overline{u}}

\newcommand{\uu}{\underline{u}}

\newcommand{\Ub}{\overline{U}}
\newcommand{\Ubb}{\overline{\overline{U}}}
\newcommand{\Uu}{\underline{U}}

\newcommand{\Vb}{\overline{V}}

\newcommand{\wb}{\overline{w}}

\newcommand{\wu}{\underline{w}}

\newcommand{\nn}{\nonumber}

\newcommand{\C}{\mathbb{C}}

\newcommand{\ord}{\mathcal{O}}

\newcommand{\pv}{\text{Painlev\'e }}

\newcommand{\pI}{\text{P$_{\text{I}}$}}
\newcommand{\pII}{\text{P$_{\text{II}}$}}
\newcommand{\pIII}{\text{P$_{\text{III}}$}}

\newcommand{\pVI}{\text{P$_{\text{VI}}$}}

\newcommand{\bs}{\backslash}
\newcommand{\cmmnt}[1]{}

\newcommand{\1}{\:\!}
\newcommand{\2}{\;\!}

\theoremstyle{definition}

\numberwithin{equation}{section}
\numberwithin{figure}{section}
\allowdisplaybreaks
\begin{document}
\title{Elliptic asymptotics in $q$-discrete Painlev\'{e} equations}
\author{Nalini Joshi}
\address{School of Mathematics and Statistics F07, The University of Sydney, NSW 2006, Australia}
\email{nalini.joshi@sydney.edu.au}
\thanks{This research was supported by an Australian Laureate Fellowship \# FL 120100094 from the Australian Research Council. }
\author{Elynor Liu}
\email{e.liu@maths.usyd.edu.au}
\address{School of Mathematics and Statistics F07, The University of Sydney, NSW 2006 Australia}
\thanks{EL's research was supported by a postgraduate research award from the University of Sydney.}
\date{}
\begin{abstract}
We study the asymptotic behaviour of two multiplicative- ($q$-) discrete Painlev\'e equations as their respective independent variable goes to infinity. It is shown that the generic asymptotic behaviours are given by elliptic functions. We extend the method of averaging to these equations to show that the energies are slowly varying. The Picard-Fuchs equation is derived for a special case of $q$-\pIII.
\end{abstract}
\maketitle
\section{Introduction}\label{sec:intro}

Despite widespread interest in the asymptotic analysis of the Painlev\'e equations, many corresponding questions on asymptotic behaviours of discrete Painlev\'e equations remain open. Asymptotic results for $q$-discrete Painlev\'e equations are particularly scarce (with only a few known cases), even though they arise in cluster algebra \cite{Okub2013} and in problems related to gap probability functions \cite{Kniz2016}. 

In this paper, we analyse the asymptotic behaviours of general solutions of $q$-discrete Painlev\'e equations and show that they are asymptotic to elliptic functions, in a way that closely resembles the behaviours of their continuous counterparts.   This study is motivated by our recent study of the elliptic asymptotics of the first to the fifth Painlev\'e equations \cite{JL2018}. We restrict our attention to the limit when the independent variable $\xi=\xi_0 q^n$ approaches infinity with $|q|>1$ and $\Re(n)>0$.

We focus on the following $q$-discrete Painlev\'e equations:
\begin{align}\label{intro:qp1}
\text{q-}\pI:&\qquad\wb\2w\2\wu=1-\frac{1}{\xi \2 w},\\
\label{intro:qp3}
\text{q-}\pIII:&\qquad\wb\2\wu=\frac{c\1d\1(w-a\1\xi)(w-b\1\xi)}{(w-c)(w-d)},
\end{align}
where $w=w(\xi)=w_n$, $\wb=w(q\xi)=w_{n+1}$, $\wu=w(\xi/q)=w_{n-1}$ with $\xi=\xi_0q^n$ for $\xi_0\in\C\backslash \{0\}$. Further information about the derivation and properties of these equations, including their respective continuum limits to the first and third Painlev\'e equations, are provided in Section \ref{s:bg}. In an instance of differing nomenclature, these equations are also referred to in the literature by the root systems characterising their initial value space and symmetry group \cite{Sakai2001}. The equation labelled $q$-$\pI$ has initial value space $A_7^{(1)}$ and symmetry group $A_1^{(1)}$, while that labelled $q$-$\pIII$ has initial value space $A_3^{(1)}$ and symmetry group $D_5^{(1)}$.

Our main results are stated as Theorems \ref{thm:qp1} and \ref{thm:qp3} in Section \ref{s:main}. We write $q=1+\frac{1}{\eta}$ and assume that $n$ lies in a domain near infinity by taking $n=\eta^2+m$ where $m$ is bounded, with $|\eta|\to\infty$ and $\Re(n)>0$.  We start by transforming variables in such a way that the equations become autonomous and of order unity to leading order as $|\eta|\to\infty$. For $q$-$\pI$, no transformation is needed, hence $u=w$. For $q$-$\pIII$ the scaling $u=e^{\frac12-\eta}w$ is applied, hence Equation \eqref{intro:qp3} becomes
\be\label{intro:qp3:u}
\ub \uu=\frac{\gm\2\de(u-\al\2\Xi)(u-\beta\1\Xi)}{(u-\gm)(u-\de)},
\ee
where $c=e^{\eta-\frac12}\gm$, and $d=e^{\eta-\frac12}\de$, and renaming the parameters $\al:=a\2\xi_0$, $\beta:=b\2\xi_0$, and $\Xi:=\frac{\xi}{\xi_0e^{\eta-\frac12}}$ in the limit $|\eta|\to\infty$, for more details see Section \ref{qp3:pre}. In the following section,  we use these transformations to set up our initial value problems and express our main results. 

We show that each leading order equation has an invariant, which enables us to identify an energy-like quantity called $E$ (defined in Equations \eqref{intro:inv:qI} and \eqref{intro:inv:qIII}), which is related to the modulus of the leading-order elliptic function. We will refer to $E$ as energy for short. Techniques from the calculus of differences are used to show that the modulation of the energy is small. In this paper we apply this approach to only two cases of $q$-discrete Painlev\'e equations, however,  given appropriate scaling, we expect our approach  to be extendable to other $q$-discrete Painlev\'e equations.

Below we state our main results in detail, provide a description of background and give an outline of the paper.

\subsection{Main result}\label{s:main}
\hfill\\
First we provide some preliminary definitions and remarks to describe the initial value problems and clarify our results. We shall assume that all the variables and parameters in Equations \eqref{intro:qp1} and \eqref{intro:qp3} are complex, and that the discrete dependent variable is at least once differentiable as a function of its independent variable. Our major results are stated as Theorems \ref{thm:qp1} and \ref{thm:qp3}. 

Recalling that  $\xi=\xi_0\1 q^{n}$, the iterations of $q$-$\pI$, $q$-$\pIII$ can be thought of as iterating $n$ on a line $\ell$ in $\C$ or iterating $\xi$ on a spiral $\Gamma\subset\C$. Given an initial point $\xi_0$ on $\Gamma$, we assume that initial values $u_0$, $u_1$ are given as analytic functions in a domain $\mathcal D_0$ containing $\xi_0$ and $q\xi_0$ as interior points.  We assume $q\in\C\bs\{0,1\}$ is such that subsequent iterations give solutions in overlapping domains with points on $\Gamma$ as interior points. Alternatively, we can regard the iteration as giving successive overlapping domains $\mathcal D$ with points on the horizontal line $\ell$ as interior points. With this in mind, we define admissible initial conditions for each respective discrete equation below.
 
\begin{figure}[H]
\includegraphics[scale=0.4]{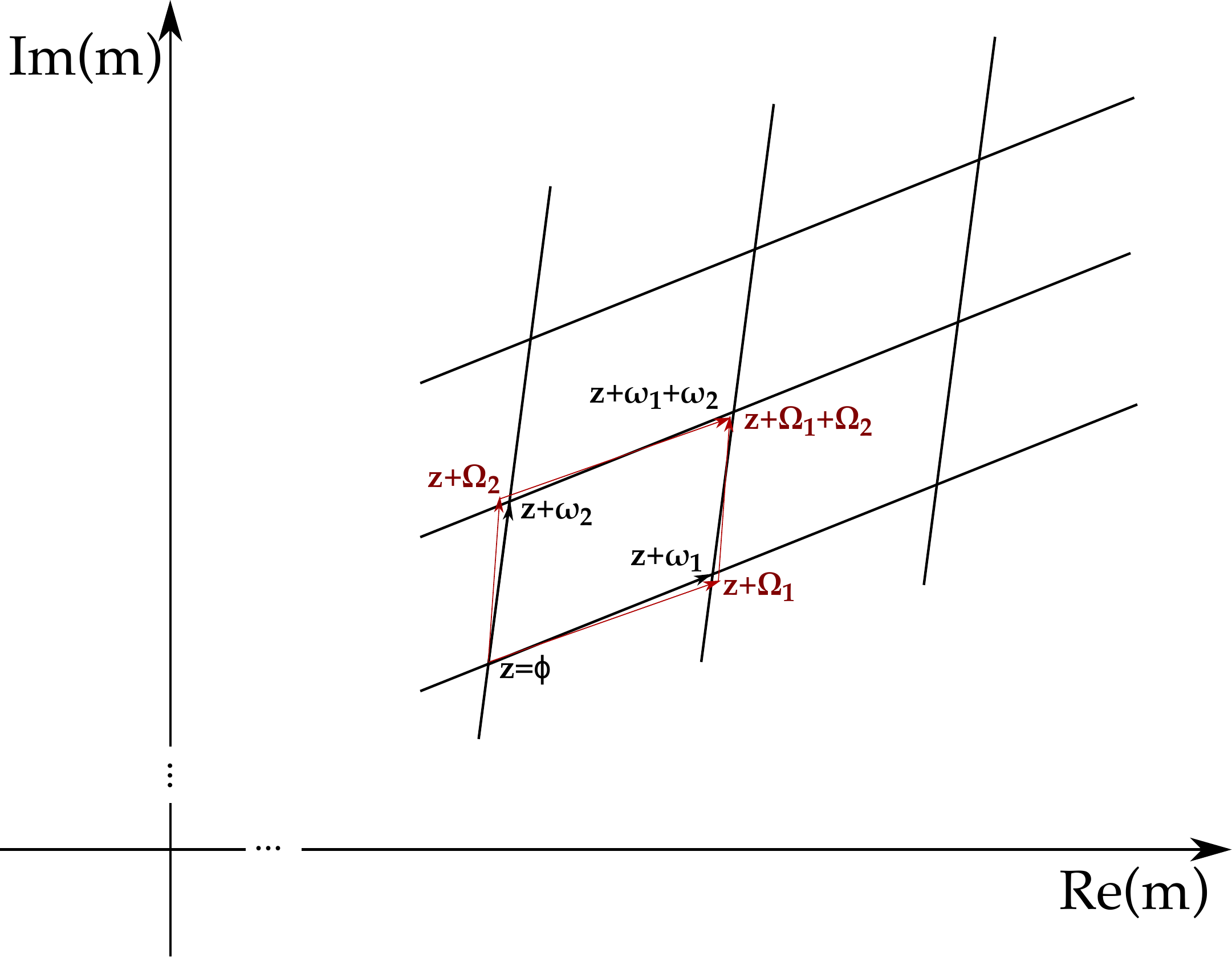}
\caption{Schematic representation of leading order periods $\om_1$ and $\om_2$, and the approximate-periods $\Om_1$ and $\Om_2$ for $u$ at $\phi$.}\label{fig:periods}
\end{figure}

{\defn\label{def2}
Given real $\delta>0$, define $\mathcal D_\de=\{|m|<\de\}$, and assume that $\phi$ and $\phi+1$ lie in $\mathcal D_\de$. For such $\phi$, assume $a$, $p$ are given initial values for a solution $u$ of $q$-$\pI$ or $q$-$\pIII$, such that 
\[
u(\phi)=a, \; u(\phi+1)=p,
\]
where $p\neq a$ and $a\neq 0$ for $q$-$\pI$ and $q$-$\pIII$, $p\neq 0$ for $q$-$\pIII$, $p\neq \frac{1}{\xi_0 q^\phi}$ for $q$-$\pI$, and $p\neq\al$, $\beta$, $\gm$, or $\de$ for $q$-$\pIII$.  We define the set of numbers $\phi$, $a$, $p$ satisfying the above conditions to be {\em admissible}.  
}
{\defn
Let $\phi$, $a$, and $p$ be admissible data. We define $\Om_i^{(J)}$ to be the distance to the next occurrence of the initial value $a$ in the direction of $\om_i^{(J)}$, the periods for the leading order elliptic behaviour,
\be\label{Omega}
u(\phi+\Om_i^{(J)})=u(\phi)=a,
\ee
for $i=1,\, 2$ and $J=\rm I, III$. We call this quantity an {\em approximate-period}. See Figure \ref{fig:periods}.}

{\remark Note that for any admissible initial values in $\mathcal D_\delta$, the mapping given by $q$-$\pI$ and $q$-$\pIII$ is analytic. Therefore, the derivative of the solution cannot be identically zero in the image domain. Since we are working in a bounded domain in $m$, without loss of generality, we assume that  $\mathcal D_\delta$ and its images do not contain a zero of the derivative of $u$. It follows that we can apply the inverse function theorem to Equation \eqref{Omega} in the resultant domain to show that $\Om_i^{(J)}$ exists. 
}

Below we state our main theorems, in which we use the dependent variable $u$ and independent variable $m$. We also drop the superscripts $\rm(I)$ and $\rm(III)$ when there is no ambiguity.
\begin{theorem}[{q-$\pI$}]\label{thm:qp1}
Let $\phi$, $a$, $p$ be admissible data and assume $u=U+s$ is a solution of $q$-$\pI$, satisfying $u(\phi)=a, u(\phi+1)=p$, where $s\ll U$ as $|\eta|\to\infty$. Define the energy
\be
E_{\rm I}\;\,:=\frac{\ub^2 u^2+\ub+u}{\ub\1u}.\label{intro:inv:qI}
\ee
Then for sufficiently large $\eta$, there exist approximate-periods, $\Om_i$, $i=1, 2$, such that $\De_{i} E_{\rm I}:=E_{\rm I}(\phi+\Om_i)-E_{\rm I}(\phi)$ take the form:
\be
\De_{i} E_{\rm I}=\frac{\zeta_0}{e^\eta}\left(\frac{\om_i}{\eta\2 U_{\phi+1} U_\phi}+\frac1\eta\sum_{k=\phi+1}^{\phi+\Om_i-1}\frac{1}{U_{k+1} U_k}+\ord\left(\frac1{\eta^2}\right)\right).
\ee
Furthermore, the approximate-periods $\Om_i$ are given by:
\be
\Om_i=\om_i-\frac{s(\phi+\om_i)}{U'(\phi)}+o(s),
\ee
where $'=\frac{d\,}{dm}$, and 
\beq
s_m&=&F_m\frac{\zeta_0}{e^\eta}\sum_{k=\phi}^{m-1}\frac{1}{F_{k+1}F_k}\left(1-\frac{U_{k+1}U_k}{U_{\phi+1}U_\phi}\right),
\eeq
where $F_m=U_{m+1}U_m^2-U_m^2U_{m-1}$.
\end{theorem}
\begin{theorem}[{q-$\pIII$}]\label{thm:qp3}
Let $\phi$, $a$, $p$ be admissible data and assume $u=U+s$ is a solution of  $q$-$\pIII$, satisfying $u(\phi)=a, u(\phi+1)=p$, where $s\ll U$ as $|\eta|\to\infty$. Define the energy $E_{\rm III}$:
\be
E_{\rm III}:=\frac{\ub^2 u^2-(\gm+\de)\2\ub u\left(\ub+u\right)+\gm\de\left(\ub^2+u^2\right)-(\al+\beta)\gm\de\left(\ub+u\right)+\al\beta\gm\de}{\ub u}.\label{intro:inv:qIII}
\ee
Then for sufficiently large $\eta$, there exist approximate-periods, $\Om_i$, $i=1,2$, such that $\De_{i} E_{\rm III}:=E_{\rm III}(\phi+\Om_i)-E_{\rm III}(\phi)$ take the form:
\begin{equation}
\begin{aligned}
\De_{i} E_{\rm III}&=\frac{(\al+\beta)\gm\de}{\eta}\left(\left(\frac{1}{U_{\phi+1}}+\frac{1}{U_{\phi}}\right)\om_i-\sum_{k=\phi+1}^{\phi+\Om_i}\left(\frac{1}{U_{k+1}}+\frac{1}{U_k}\right)\right)\\
&\qquad -\frac{2\al\beta\gm\de}{\eta}\frac{\om_i}{U_{\phi+1}U_{\phi}}%
+\frac{2\al\beta\gm\de}{\eta}\sum_{k=\phi}^{\phi+\Om_i}\frac{1}{U_{k+1} U_k}%
+\ord\left(\frac{1}{\eta^{2}}\right).
\end{aligned}
\end{equation}
Furthermore, the approximate-periods $\Om_i$ are given by:
\be
\Om_i=\om_i-\frac{s(\phi+\om_i)}{U'(\phi)}+o(s),
\ee
where $'=\frac{d\,}{dm}$, and 
\begin{equation}
\begin{aligned}
s_m&=F_m\!\sum_{l=\phi}^{m-1}\!\!\left(\!E_1\frac{U_{l+1}U_l}{F_{l+1}F_l}%
+\frac{(\al+\beta)\gm\de}{3\eta}\frac{(3l+4)(U_{l+1}+U_l)}{F_{l+1}F_l}\right.\\
&\hspace{20mm}-\frac{2\al\beta\gm\de}{3\eta}\frac{3l+4}{F_{l+1} F_l}%
+\frac{2\al\beta\gm\de}{\eta}\frac{U_{l+1}U_l}{F_{l+1}F_l}\!\sum_{k=\phi}^{l}\frac{1}{U_{k+1}U_k}\\%
&\hspace{30mm}\left.-\frac{(\al+\beta)\gm\de}{\eta}\frac{U_{l+1}U_l}{F_{l+1}F_l}\!\sum_{k=\phi}^{l}\!\!\left(\!\frac1{U_{k+1}}+\frac1{U_k}\right)\!\!\right)\!,\label{qp3:s:leadsol}
\end{aligned}
\end{equation}
where $F_m=(U_{m+1}-U_{m-1})(U_m-\gm)(U_m-\de)$.
\end{theorem}
We will use the following lemmas in the analysis of both $q$-$\pI$ and $q$-$\pIII$.
{\lemma\label{xi:exp} Suppose that $\eta$ is large, then for $\xi=\xi_0q^n$ ($\xi_0\in\C\backslash\{0\}$), where $|q|>1$ and $\Re(n)>0$ with $q=1+\frac{1}{\eta}$ and $n=\eta^2+m$ where $m$ is bounded, $\xi$ takes the following expansion in the limit $|\eta|\to\infty$:
\be\label{xi_exp}
\xi=\xi_0e^{-\frac{1}{2}}e^{\eta}\left(1+\frac{3m+1}{3\eta}+\frac{18m^2-6m-7}{36\eta^2}+\ord\left(\frac{1}{\eta^3}\right)\right).
\ee}
\begin{proof}
The proof follows by direct computation.
\end{proof}
{\lemma\label{ubexpan:2nd} Suppose $s(m)=\lam \2 G(m)\sum_{k=\phi}^{m-1}R(k)$, where $\lam$ is independent of $m$, and $G$ is periodic in $\om$, then 
\be
s(m+\om+1)-s(m+\om)\frac{G(m+\om+1)}{G(m+\om)}=\lam\2G(m+1)R(m+\om).
\ee}
\begin{proof} This can be shown by direct expansion of the sum $s(m)$.
\end{proof}
{\lemma\label{ubexpan} The expression $u(\phi+\Om+1)$ has the following expansion:
\beqn
u(\phi+\Om+1)&=&U(\phi+1)+\lam\2G(\phi+1)R(\phi+\om)+\ord\left(s^2\right),
\eeqn
where $u=U+s$, $s\ll U$, in the limit where the independent variable goes to infinity. Furthermore, $\Om$ is of the following form:
\be
\Om=\om+\left.\left(-\frac{s}{U'}+\frac{ss'}{(U')^2}-\frac{s^2U'' }{2(U')^3}\right)\right|_{n=\phi+\om}+\ord(s^3).
\ee
\begin{proof}
See Appendix \ref{ubexpan:pf}.
\end{proof}
Proofs of Theorems \ref{thm:qp1} and \ref{thm:qp3} can be found in Section \ref{sec:qp1} and Section \ref{sec:qp3} respectively.
\subsection{Background}\label{s:bg}
Discrete Painlev\'e equations have a long history, beginning with the theory of orthogonal polynomials \cite{Sho1939}. In particular, Shohat derived a nonlinear recurrence relation that was  later recognised as a discrete Painlev\'e equation by Fokas \emph{et al} \cite{FIK1992}. Around the same time, such discrete equations appeared in models of quantum gravity \cite{GM1990,PS1990}.

Our equations of interest, $q$-$\pI$ and $q$-$\pIII$, were derived and studied by Ramani \emph{et al} \cite{RG1996, RGH1991}. These authors also showed that the conventional form of Equation \eqref{intro:qp1}\footnote{See \cite{Josh2015} for the transformation of $q$-$\pI$ to its conventional form.} has a continuum limit to the first Painlev\'e equation $y_{tt}=6y^2+t$, while Equation \eqref{intro:qp3} has a continuum limit to the third Painlev\'e equation \cite{RGH1991}
 \ben
y_{tt}=\frac{y_t^2}{y}+e^t(\al\2y^2+\beta)+e^{2t}\frac{\gm y^4+\de}{y},
\een
which is a transformed version of the conventional form of $\pIII$ (see \cite[Chapter 14]{Ince1956} for the transformation).

These equations also arise in other fields of mathematical physics. Equation \eqref{intro:qp1} arises in cluster algebras as an equation related to the non-autonomous version of Somos-4 \cite{Okub2013}. Recently, the asymmetric version of Equation \eqref{intro:qp3}, known as $q$-$\pVI$, has been studied from the point of view of conformal block theory \cite{JNS2017}. 

These perspectives lead to a natural question about the behaviours of solutions. Nishioka \cite{Nish2010} showed that the solutions of $q$-$\pI$ are highly transcendental functions in the sense that they cannot be expressed in terms of any earlier known functions or solutions of linear equations with such functions as coefficients. A similar result is believed to be true for $q$-$\pIII$, since $q$-$\pI$ arises as a limiting equation from $q$-$\pIII$. 

However, there appears to be no explicit information about such transcendental solutions, except for studies of asymptotic behaviours, which have been developed for a few cases of $q$-discrete Painlev\'e equations in the complex plane \cite{Mano2010,RofThesis,JRo2016,Josh2015,Ohya2010}. In these studies, solutions asymptotic to power series expansions in the independent variable were studied for $q$-$\pVI$, $q$-P$(A_1)$, and $q$-$\pI$. For $q$-$\pI$, unstable solutions termed \emph{quicksilver} solutions were identified \cite{Josh2015}. The geometric description of the space of initial values in the asymptotic limit $|\xi|\to\infty$ was given in \cite{JL2016}. In the latter study, the invariant for the autonomous leading-order equation was also considered. 
This contrasts with the studies of classical Painlev\'e equations, where general solutions asymptotic to elliptic functions are well known \cite{JL2018}.  

Elliptic-function-type asymptotic behaviours have been studied in the case of the additive first discrete Painlev\'e equation, d-$\pI$ \cite{Vere1996, Josh1997}. Moreover, the Stokes phenomenon has been studied for the solution behaviours of additive discrete $\pI$ and $\pII$ equations \cite{JL2015, JLL2017}.

A general theory concerning the asymptotic analysis of difference equations can be found in \cite{HS1964,HS1965}. Stokes phenomena in difference equations were also studied by Dingle \cite{Ding1973}. However, many of the techniques developed to study asymptotic behaviours of ODEs have not yet been extended to difference equations. In particular the method of averaging has not been applied, prior to the present work, to study $q$-difference equations. The standard method of averaging is an approximation method often used for weakly nonlinear dynamical systems. It allows one to infer properties of the original dynamic system by understanding the dynamics of the averaged system. Although the standard averaging theorem addresses continuous functions in a real domain, it can be applied to any continuous almost periodic function in the complex plane. These extensions have been studied and implemented in \cite{JoshiThesis, Mahm1998,Hust1970, HB2009, SA2003}. For its statement, proof and more details, we refer to \cite{Ver1996, SVM2007}. As in the continuous setting, we find the method of averaging to be well-suited for deriving the asymptotic behaviour of almost periodic discrete equations. 

\subsection{Outline of the paper} 
In Section \ref{sec:qp1}, we give a detailed study of $q$-$\pI$. The analysis is divided into three subsections which respectively include the leading-order analysis, next-to-leading order analysis and analysis of the variation of  the invariant. In Section \ref{sec:qp3}, we study $q$-$\pIII$ in a similar way, with an extra subsection that treats a special case. This special case allows an explicit formulation of the leading order solution, its periods, and its Picard-Fuchs equation.

\section{The first $q$-discrete \pv equation}\label{sec:qp1}
In this section, we perform our asymptotic analysis on $q$-\pI. No scaling is needed in the limit $|\xi|\to\infty$, but to remain consistent with the notation used for $q$-$\pIII$ later we rewrite the equation in the form:
\be\label{qp1:u}
\ub\2u\2\uu=1-\frac{1}{\xi \2 u}.
\ee
We start by analysing the leading-order behaviour in Section \ref{qp1:sec:lead}. The next-to-leading-order behaviour is then calculated in Section \ref{qp1:sec:small}. In Section \ref{qp1:sec:avg} we prove that the modulation of the energy varies slowly as $|\xi| \rightarrow \infty$. The subscript $\rm I$ denoting $q$-\pI will be dropped in this section for simplicity.
\subsection{Leading-order analysis}\label{qp1:sec:lead}
In the limit $|\xi|\to\infty$, Equation \eqref{qp1:u} becomes
\be\label{qp1:asym}
\ub u\uu\sim1,
\ee
to leading order. We define the leading-order $U$ as the solution of  
\be\label{qp1:leada}
\Ub U\Uu=1,
\ee
which can be integrated (actually summed) once to give 
\be
E_0=\frac{\Ub^2 U^2+\Ub+U}{\Ub U},
\ee
where $E_0$ is a constant.  
It can be readily shown that $U$ is periodic with period 3. 

This leads to the asymptotic expansion: $u=U+s$, where $s\ll U$ as $|\xi|\to\infty$. 
Motivated by this, we define an energy-like quantity, which is constant to leading-order as shown in the following lemma.
{\lemma \label{qp1:E} Define
\be\label{qp1:inv}
E:=\frac{\ub^2u^2+\ub+u}{\ub\2u},
\ee
where $u$ satisfies Equation \eqref{qp1:u}. Then, $E$ is constant to leading order in the limit $|\xi|\to\infty$.}
\begin{proof} The proof follows by direct calculation using the integration factor $\frac{1}{\uu}-\frac{1}{\ub}$.
\end{proof}
{\cor\label{qp1:cor:E} The change of $E$, $\De E=E(m+1)-E(m)$, is given by
\be\label{qp1:E:cor}
\De E=\left(\frac{1}{\ub}-\frac{1}{\uu}\right)\frac{1}{\xi \2 u}.
\ee
}
The corollary follows directly from the proof of the lemma. 

The expansion of $\xi$ as $|\eta|\to\infty$ is given by Lemma \ref{xi:exp} and, therefore, we have the following result.
{\lemma Equation \eqref{qp1:u} has the following expansion as $|\eta|\to\infty$:
\be\label{qp1:expan}
\ub u\uu=1-\frac{\zeta_0}{e^{\eta} \1 u}\left(1-\frac{3m+1}{3\eta}+\frac{18m^2+30m+11}{36\eta^2}+\ord\left(\frac{1}{\eta^3}\right)\right),
\ee
where $\zeta_0:=\frac{e^{1/2}}{\xi_0}$.}
{\cor \label{qp1:cor:E:RHS} Corollary \ref{qp1:cor:E} and Lemma \ref{xi:exp} imply that $E$ has the following expansion
\be\label{qp1:E:RHS}
E=E_c+\frac{\zeta_0}{e^\eta}\left(\frac{1}{\ub\1u}-\frac{3m+4}{3\eta\ub u}+\sum_{k=\phi}^m\frac{1}{\eta u_{k+1}u_k}+\ord\left(\frac{1}{\eta^2}\right)\right),
\ee
where $E_c$ is the integration constant in $m$.}\\
Note that $E_c$ is constant as a function of $m$, but may have an expansion in powers of $\eta$. Using $u=U+s$, we can expand Equation \eqref{qp1:u} in powers of $\eta$. This leads to the following lemma.
{\lemma For admissible initial conditions, Equation \eqref{qp1:u} becomes
\beq
&&\Ub U \Uu=1,\label{qp1:lead}\\
&&\Ub\Uu s+\Ub U\su+U\Uu \sbar+\Ub s\su+\Uu \sbar s+ U \sbar\su+\sbar s \su=-\frac{1}{\xi}\sum_{k=0}^{\infty}\frac{(-s)^k}{U^{k+1}}.\label{qp1:small}
\eeq
Here, $U$ and its iterations are of order unity, while $s$ and its iterations are of order $e^{-\eta}$.
}
\begin{proof}%
Equations \eqref{qp1:lead} and \eqref{qp1:small} can be generated by direct substitution of $u=U+s$ into \eqref{qp1:u} and equating terms of equal order. The admissible initial conditions are used to allow \eqref{qp1:small} to remain bounded. $U$ and its iterations are solved by Weierstrass $\wp$-functions (see Lemma \ref{qp1:wpsoln} below), and hence are of order unity. That $s$ and its iterations are of order $e^{-\eta}$ is shown by using dominant balance analysis on \eqref{qp1:small}.
\end{proof}
{\cor In terms of the asymptotic expansion, the initial values now require:
\be\label{qp1:IV}
\bc
U(\phi)=a\\
s(\phi)=0
\ec\!\!\!\!,\qquad
\bc
U(\phi+1)=p\\
s(\phi+1)=0
\ec\!\!\!\!.
\ee}
{\lemma\label{qp1:wpsoln} The leading-order equation \eqref{qp1:lead} satisfied by $U$ is solved by a Weierstrass $\wp$-function.}
\begin{proof}
Equation \eqref{qp1:lead} can be summed once to
\be\label{qp1:E0}
\Ub^2 U^2 -E_0\Ub U+ \Ub + U=0,
\ee
where $E_0$ is the integration constant. The transformations $U=x+\frac{E_0^2}{12}$ and $\Ub=y+\frac{E_0^2}{12}$ convert \eqref{qp1:E0} into the following biquadratic polynomial,
\be\label{qp1:inv:wp}
x^2y^2+\frac{E_0^2}{6}xy(x+y)+\frac{E_0^4}{144}(x^2+y^2)+\gm\1 xy+\de(x+y)+\ep=0,
\ee
where
\beqn
\gm&=&\frac{E_0^4}{36} -E_0,\\
\de&=&\frac{E_0^6}{864}-\frac{E_0^3}{12}+1,\\
\ep&=&\frac{E_0^8}{20736}-\frac{E_0^5}{144}+\frac{E_0^2}{6}.
\eeqn
Equation \eqref{qp1:inv:wp} can be parametrised by $x=\wp(z)$ and $y=\wp(z\pm\rho)$, where 
\begin{subequations}
\begin{align}
&\rho=\wp^{-1}\left(-\frac{E_0^2}{12}\right),\\
&g_2=\frac{E_0^4}{12}-2E_0,\quad g_3=\frac{E_0^6}{216}-\frac{E_0^3}{6}+1,\\
&m=\frac{z}{\rho},\quad m+1=\frac{z+\rho}{\rho}.
\end{align}
\end{subequations}
We note here that $\rho$ is not unique since the inverse of the Weierstrass $\wp$-function gives two values in any given period parallelogram. Therefore the leading order solution $U$ takes the following form:
\beqn
U(m)&=&\wp(z)+\frac{E_0^2}{12}.
\eeqn
The forward iteration is $U(m+1)=\wp(z\pm\rho)+\frac{E_0^2}{12}$. The sign of $\rho$ which determines the step in $m$ depends on the initial values. 
\end{proof}
The above results in the differential equation satisfied by $U$,  which in turn allows the expression of its period integrals.
{\cor From Lemma \ref{qp1:wpsoln}, we can conclude that $U$ satisfies the following differential equation with respect to $z$,
\be\label{qp1:Uzeq}
U_z^2=4U^3-E_0^2U^2+2E_0 U-1.
\ee
Therefore, we define the periods by:
\be\label{qp1:om}
\om_j=\oint_{C_j}\frac{dU}{\sqrt{4U^3-E_0^2U^2+2E_0 U-1}},
\ee
where $j=1,2$ and $C_j$ are linearly independent contours.}
{\remark Equation \eqref{qp1:E0} has $4$ degenerate points at $E_0=3$, $3(-1)^{2/3}$, $3(-1)^{4/3}$, and $\infty$. This number can be reduced to $2$ under the identification $K^{1/3}=E_0$. In this case $K=27$ and $\infty$ are the degenerate points.}

The elliptic integral \eqref{qp1:om} has three branch points in the finite plane and one at infinity, which results in two linearly independent periods. At the degenerate points, the integrand in \eqref{qp1:om} develops a simple pole and it is no longer possible to define two linearly independent contours. Hence $U$ becomes singly periodic in these cases.  We note that the sequence $U$ and its analytic continuation in terms of the Weirestrass $\wp$ function retain the periodicity of 3 implied by Equation \eqref{qp1:leada}. 

{\cor The following properties can be derived from \eqref{qp1:E0}:
\beq
&&{\rm i).}\quad\frac{d\2\Ub}{d\2U}=-\frac{\Ub^2(\Ub U^2-1)}{U^2(\Ub^2U-1)},\label{qp1:dUbdU}\\
&&{\rm ii).}\quad \Ub=\frac{E_0 U-1\pm\sqrt{-4U^3+E_0^2U^2-2E_0U+1}}{2U^2}\label{qp1:Ub:sol},
\eeq
where in deriving \eqref{qp1:dUbdU} we assumed that $E_0$ is not a function of $U$ and $\Ub$.
}

\subsection{Next-to-leading order analysis}\label{qp1:sec:small}
{\lemma \label{qp1:sl} Suppose that $s_l$ is the leading order behaviour of $s$, where $s_l$ satisfies the limiting equation \eqref{qp1:small} as $|\eta|\to\infty$, then $s_l$ is given by
\be\label{qp1:sl:eq}
s_l(m)=F_m\frac{\zeta_0}{e^\eta}\sum_{k=\phi}^{m-1}\frac{1}{F_{k+1}F_k}\left(1-\frac{U_{k+1}U_k}{U_{\phi+1}U_\phi}\right),
\ee
where $s_l(\phi):=0$ and $F:=\Ub U^2-U^2\Uu$.}
\begin{proof}
We use the asymptotic expansion $u=U+s$ to expand \eqref{qp1:inv} and \eqref{qp1:E:RHS}. Equating  the order of $s_l$ (i.e. order of $e^{-\eta}$) terms gives the following equation:
\be\label{qp1:sleq}
\frac{\Ub^2U^2-U}{\Ub}\2\sbar_l+\frac{\Ub^2U^2-\Ub}{U}\2 s_l =\frac{\zeta_0}{e^\eta}+E_1\Ub U,
\ee
where $E_1$ is derived from the initial conditions
\be
E_1=-\frac{\zeta_0}{e^\eta\1U_{\phi+1}U_\phi}.
\ee
Equation \eqref{qp1:sleq} can be integrated by recognising that two of its coefficients are exact differences, that is, we have
\beq
\frac{\Ub^2U^2-U}{\Ub}&=&\frac{\frac{\Ub U}{\Uu}-U}{\Ub}=\frac{\Ub U-U\Uu}{\Ub\Uu}=\Ub U^2-U^2\Uu,\label{qp1:sl:1}\\
\noalign{\textrm{and}}\nonumber\\
\frac{\Ub^2U^2-\Ub}{U}&=&\frac{\frac{\Ub U}{\Ubb}-\Ub}{U}=\frac{\Ub U - \Ubb\Ub}{\Ubb U}=-(\Ubb\2 \Ub^2-\Ub^2 U),\label{qp1:sl:2}
\eeq
where we have used Equation \eqref{qp1:lead} in Equation \eqref{qp1:sl:1} and \eqref{qp1:sl:2} uses a forward iteration of \eqref{qp1:lead}. Using the definition of $F$ and dividing \eqref{qp1:sleq} by $\Fb\1F$ give 
\beqn
&&\frac{\sbar_l}{\Fb}-\frac{s_l}{F}=\frac{\zeta_0}{e^\eta\1\Fb\1F}\left(1-\frac{\Ub U}{U(\phi+1)U(\phi)}\right).
\eeqn
Summing the above yields the desired result, where the integration constant turns out to be zero. Note that for $s_l$ to be well-defined, we have assumed (w.l.o.g.) that $F$ is non-zero in the domain $\mathcal D_\delta$.
\end{proof}
{\remark $F$ can also be expressed as:
\be
F(U,E_0)=\pm\sqrt{-4U^3+E_0^2U^2-2E_0U+1},
\ee
by using \eqref{qp1:Ub:sol}.}
Notice that  $F$ and each component of $s_l$ are periodic in $\om$. Moreover, we  note that $\frac{d\Ub}{dU}=\frac{\Fb}{F}$.
{\cor\label{qp1:approxperiod} The approximate-period $\Om$  in the direction of $\om$ has the following expansion:
\be\label{qp1:Om}
\Om=\om-\frac{s_l(\phi+\om)}{U'(\phi)}+\ord\left(\frac{1}{\eta e^{\eta}},\frac{1}{e^{2\eta}}\right),
\ee
where $s_l$ can be represented by \eqref{qp1:sl:eq} and $U'=\frac{d\1U}{d\1m}$, with
\be
\left(\frac{d\1U}{d\1m}\right)^2=\rho^2(4U^3-E_0^2U^2+2E_0U-1),
\ee
which follows by transforming \eqref{qp1:Uzeq} into $m$.}
\subsection{Evolution of the energy}\label{qp1:sec:avg}
In this section we derive the local change of the energy across an approximate-period. 

{\lemma\label{qp1:ubexp} In the case of $q$-\pI, $\lam=\frac{\zeta_0}{e^\eta}$, $G=F$, and $R(k)=\frac{1}{F_{k+1}F_k}\left(1-\frac{U_{k+1}U_k}{U_{\phi+1}U_\phi}\right)$, hence
\be
u(\phi+\Om+1)=U(\phi+1)+\ord\left(\frac{1}{\eta \2 e^{\eta}}\right).
\ee
}
\begin{proof}
This can be shown by using Lemmas \ref{ubexpan:2nd} and \ref{ubexpan}.
\end{proof}
We are now in a position to evaluate the change of the energy across an approximate-period. In the proof below, we use \eqref{qp1:E:RHS} to represent $E$.
\hfill\\

{\prop\label{qp1:avg}The evolution of the energy $E$ of $q$-$\pI$ is given by
\be\label{qp1:avg:eq} 
E(\Phi+\Om)-E(\Phi)=\frac{\zeta_0}{e^\eta}\left(\frac{\om}{\eta\2 U_{\phi+1} U_\phi}+\frac1\eta\sum_{k=\phi+1}^{\phi+\Om-1}\frac{1}{U_{k+1} U_k}+\ord\left(\frac1{\eta^2}\right)\right).
\ee}
\begin{proof}
We start with the expansion \eqref{qp1:E:RHS} for $E$. Consider the difference:
\be\label{qp1:avg:1}
\frac{(E(\Phi+\Om)-E(\Phi))e^\eta}{\zeta_0}=\left.\left(\frac{1}{\ub\1u}-\frac{3m+4}{3\eta\ub u}+\sum_{k=\phi}^{m}\frac{1}{\eta u_{k+1}u_k}+\ord\left(\frac{1}{\eta^3}\right)\right)\right|_{m=\phi}^{m=\phi+\Om}.
\ee
Consider the first term from \eqref{qp1:avg:1}, which can be reduced by  using Corollary \ref{qp1:ubexp} to give
\beq
\frac{1}{u(\phi+\Om+1)\1u(\phi+\Om)}-\frac{1}{u(\phi+1)\1u(\phi)}
&=&\ord\left(\frac{1}{\eta\2e^{\eta}}\right),
\eeq
where we have also used  property \eqref{qp1:dUbdU}. The second term from \eqref{qp1:avg:1} can be simplified into:
\beq
\frac{3\1(\phi+\Om)+4}{3\2\eta\2 u(\phi+\Om+1)\1 u(\phi+\Om)}-\frac{3\1\phi+4}{3\2\eta\2 u(\phi+1)\1 u(\phi)}=\frac{\Om}{\eta \2 U_{\phi+1}U_\phi}+\ord\left(\frac{1}{\eta\2e^{\eta}}\right).
\eeq
We then replace $\Om$ in the above equation by using \eqref{qp1:Om}. The third term can be evaluated by recognising that it is a telescoping sum. Combining the results gives the desired result. 
\end{proof}
This concludes our proof of Theorem \ref{thm:qp1}.
\section{The third $q$-discrete \pv equation}\label{sec:qp3}
In this section, we study the generic behaviour of $q$-$\pIII$ in the limit $\xi$ goes to infinity. We focus on deriving the leading and next-to-leading order behaviour. We then show that the energy is slowly varying. A special case is considered in order to find the explicit formulation of the Picard-Fuchs equation for the period integral of the leading-order behaviour. Again, we drop the subscript denoting $q$-$\pIII$.
\subsection{Leading-order analysis}\label{qp3:pre}
First we transform Equation \eqref{intro:qp3} into a form in which the asymptotic behaviour in the limit $|\xi|\to\infty$ can be readily studied.
{\lemma Using the result of Lemma \ref{xi:exp} and under the same conditions, the scaling: 
\ben
w=e^{\eta-\frac12} u, c=e^{\eta-\frac12}\gm, \text{ and } d=e^{\eta-\frac12}\de,
\een 
transforms \eqref{intro:qp3} into the following form,
\be\label{qp3:u}
\ub \uu=\frac{\gm\2\de(u-\al\2\Xi)(u-\beta\1\Xi)}{(u-\gm)(u-\de)},
\ee
where $\al:=a\2\xi_0$, $\beta:=b\2\xi_0$, and $\Xi:=\frac{\xi}{\xi_0e^{\eta-\frac12}}$, for $|\eta| \to\infty$.
}
\begin{proof}%
Lemma \ref{xi:exp} can be used to show that the controlling factor of $\xi$ is $e^\eta$ which allows us to postulate the following ansatz:
\be
w=e^{\rho\2\eta}u,\;a=e^{\sig\2\eta}\al_0,\; b=e^{\sig\2\eta}\beta_0,\; c=e^{\tht\2\eta}\gm,\; d=e^{\tht\2\eta}\de.
\ee
The method of dominant balance is used to solve for $\rho$, $\sig$, and $\tht$ under the conditions that the transformations are not singular. The resulting limiting equation is of non-degenerative QRT type in general, and a maximal number of terms is maintained. Under these conditions, we have:
\be
\rho=1,\; \sig=0,\; \text{ and } \tht=1.
\ee
This allows us to peel off the dominant behaviour of $q$-$\pIII$ and gives the resulting equation in the stated Lemma. New parameters are defined so that the leading behaviour of $\Xi$ is $1$:
\be
\Xi=1+\frac{3m+1}{3\eta}+\frac{18m^2-6m-7}{36\eta^2}+\ord\left(\frac{1}{\eta^3}\right).
\ee
\end{proof}%
{\remark In the rest of this section, the notation $\Xi_1:=\Xi-1$ is often used, where $\Xi_1$ represents the small perturbation terms.}

In the limit $|\eta|\to\infty$, Equation \eqref{qp3:u} becomes:
\be\label{qp3:asym}
\ub \uu\sim\frac{\gm\2\de(u-\al)(u-\beta)}{(u-\gm)(u-\de)},
\ee
to leading order. We define the leading-order $U$ as the solution of
\be
\Ub \Uu=\frac{\gm\2\de(U-\al)(U-\beta)}{(U-\gm)(U-\de)},
\ee
which can be integrated (actually summed) once to give
\be\label{qp3:E0}
E_0=\frac{\Ub^2 U^2-(\gm+\de)\2\Ub U\left(\Ub+U\right)+\gm\de\left(\Ub^2+U^2\right)-(\al+\beta)\gm\de\left(\Ub+U\right)+\al\beta\gm\de}{\Ub U},
\ee
where $E_0$ is a constant, also known as the invariant.  This leads to the asymptotic expansion: $u=U+s$, where $s\ll U$ as $|\eta|\to\infty$. Motivated by this, we define the energy in the following lemma.
{\lemma\label{qp3:E} Define
\be\label{qp3:inv}
E:=\frac{\ub^2 u^2-(\gm+\de)\2\ub u\left(\ub+u\right)+\gm\de\left(\ub^2+u^2\right)-(\al+\beta)\gm\de\left(\ub+u\right)+\al\beta\gm\de}{\ub u},
\ee
where $u$ satisfies Equation \eqref{qp3:u}. Then, $E$ is constant to leading order in the limit $|\xi|\to\infty$.}
\begin{proof}
This desired result yields after multiplying both sides of Equation \eqref{qp3:u} by $(u-\gm)(u-\de)$, then using the integration factor $\frac{1}{u\uu}-\frac{1}{\ub u}$. 
\end{proof}
{\cor The change of $E$, $\De E=E(m+1)-E(m)$, is given by
\be \label{qp3:corr}
\De E=\left(\frac{1}{\ub u}-\frac{1}{u\uu}\right)\left[(\al+\beta)\gm\1\de\1u\1\Xi_1-2\al\1\beta\1\gm\1\de\1\Xi_1-\al\1\beta\1\gm\1\de\1\Xi_1^2\right].
\ee
Equivalently, $E$ can also be expanded as
\beq\label{qp3:E:RHS}
E&=&E_c+\frac{(\al+\beta)\gm\de}{3\eta}\left(\frac{3m+4}{\ub}+\frac{3m+4}{u}\right)%
-\frac{(\al+\beta)\gm\de}{\eta}\sum_{k=\phi}^{m}\left(\frac1{u_{k+1}}+\frac1{u_k}\right)\nn\\%
&&\hspace{17mm}-\frac{2\al\beta\gm\de}{3\eta}\frac{3m+4}{\ub u}%
+\frac{2\al\beta\gm\de}{\eta}\sum_{k=\phi}^{m}\frac{1}{u_{k+1}u_k}+\ord\left(\frac1{\eta^2}\right),
\eeq
where $E_c$ is the integration constant in $m$.}
\lemma\label{qp3:asym} For admissible initial values, Equation \eqref{qp3:u} becomes
\begin{subequations}
\begin{align}\label{qp3:lead}
&\Ub\Uu=\frac{\gm\de\left(U-\al\right)\left(U-\beta\right)}{(U-\gm)(U-\de)},\\
&\sbar\Uu+\su\Ub+\sbar\su=\frac{H(U,s,\Xi_1)}{(U-\gm)(U-\de)}\sum_{k=0}^{\infty}\frac{\left[-(2U-\gm-\de)s-s^2\right]^k}{(U-\gm)^k(U-\de)^k},\label{qp3:s:eq}
\end{align}
\end{subequations}
where
\beq
H(U,s,\Xi_1)&=&\gm\de(2U-\al-\beta)s-(\al+\beta)\gm\de s-(\al+\beta)\gm\de U\2\Xi_1+2\al\beta\gm\de\2\Xi_1\nn\\
&&\hspace{45mm}+\al\beta\gm\de\2\Xi_1^2+\gm\de s^2.
\eeq
$U$ and its iterations are of order unity, while $s$ and its iterations are of order $\frac1\eta$.
\begin{proof}%
The decoupled equations manifest by substituting the asymptotic expansion $u=U+s$ into Equation \eqref{qp3:u} and equating the same order terms, where $\Xi_1=\ord\left(\frac1\eta\right)$ and $\Xi_1^2=\ord\left(\frac1{\eta^2}\right)$. The admissible initial conditions are imposed so that Equation \eqref{qp3:s:eq} remains bounded. Equation \eqref{qp3:lead} is an autonomous integrable equation which can be parametrised by Jacobi $\sn$-function \cite{Baxt1982}. This implies that $U$ and its iterations are of order unity. The order of $s$ and its iterations is derived from a dominant balance analysis on \eqref{qp3:s:eq}. 
\end{proof}%
{\cor Under the asymptotic expansion, the initial values now require
\be\label{qp3:IV}
\bc
U(\phi)=a\\
s(\phi)=0
\ec\!\!\!\!,\qquad
\bc
U(\phi+1)=p\\
s(\phi+1)=0
\ec\!\!\!\!.
\ee}
{\lemma\label{qp3:sol} Equation \eqref{qp3:lead} can be parametrised by elliptic functions.}
\begin{proof}
The first integral of Equation \eqref{qp3:lead} is Equation \eqref{qp3:E0}, which is parametrised by Jacobi $\sn$-function in general, see \cite{Baxt1982}. 
\end{proof}
We note that in general Equation \eqref{qp3:E0} has ten singularities counting multiplicity. These points are the degenerate points of $U$, this means that at these points $U$ becomes singly-periodic. The parametrisation is considered an analytic continuation of $U$, hence we can derive the differential equation $U$ satisfies from this parametrisation. In turn we can also define the period-integrals of $U$. For conciseness, we show these properties explicitly only for a special case in Section \ref{qp3:spe1}.

\subsection{Next-to-leading order analysis}
{\lemma Suppose that $s_l$ is the leading-order behaviour of $s$ which solves \eqref{qp3:s:eq} in the limit $|\eta|\to\infty$, then $s_l$ takes the following form:
\beq
s_l(m)&\hspace{-3mm}=&\hspace{-3mm} F(m)\!\sum_{l=\phi}^{m-1}\!\!\left(\!E_1\frac{U_{l+1}U_l}{F_{l+1}F_l}-\frac{2\al\beta\gm\de}{3\eta}\frac{3l+4}{F_{l+1} F_l}\right.\nonumber \\
&& \phantom{F(m)\!\sum_{l=\phi}^{m-1}\!\!\!E_1}
+\frac{2\al\beta\gm\de}{\eta}\frac{U_{l+1}U_l}{F_{l+1}F_l}\!\sum_{k=\phi}^{l}\frac{1}{U_{k+1}U_k} +%
\frac{(\al+\beta)\gm\de}{3\eta}\frac{(3l+4)(U_{l+1}+U_l)}{F_{l+1}F_l}\nn\\%
&&  \phantom{F(m)\!\sum_{l=\phi}^{m-1}\!\!\!E_1}
\left.-\frac{(\al+\beta)\gm\de}{\eta}\frac{U_{l+1}U_l}{F_{l+1}F_l}\!\sum_{k=\phi}^{l}\!\!\left(\!\frac1{U_{k+1}}+\frac1{U_k}\right)\!\!\right)\!\!+\!\ord\left(\!\frac1{\eta^2}\!\right),\label{qp3:s:leadsol}
\eeq
where $s_l(\phi):=0$, $F$ is defined by 
\be\label{qp3:F}
F:=(\Ub-\Uu)(U-\gm)(U-\de),
\ee
and 
\be
E_1=\frac{2\al\beta\gm\de(3\phi+1)}{3\eta\1U(\phi+1)U(\phi)}-\frac{(\al+\beta)\gm\de}{3\eta}\left(\frac{3\phi+1}{U(\phi+1)}+\frac{3\phi+1}{U(\phi)}\right).
\ee}
\begin{proof}
By equating terms of order $1/\eta$ from Lemmas  \ref{qp3:E} and \ref{qp3:asym} after substituting the asymptotic expansion $u=U+s$, we find:
\beq
&&\sbar_l\frac{\Ub^2U^2-(\gm+\de)\Ub^2U+\gm\de\Ub^2-\gm\de U^2+(\al+\beta)\gm\de U-\al\beta\gm\de}{\Ub^2 U}\nn\\
&&\hspace{4mm} +s_l\frac{\Ub^2U^2-(\gm+\de)\Ub U^2-\gm\de\Ub^2+\gm\de U^2+(\al+\beta)\gm\de\Ub-\al\beta\gm\de}{\Ub U^2}\nn\\
&&\hspace{6mm}= E_1+\frac{(\al+\beta)\gm\de}{3\eta}\left(\frac{3m+4}{\Ub}+\frac{3m+4}{U}\right)%
-\frac{(\al+\beta)\gm\de}{\eta}\sum_{k=\phi}^{m}\left(\frac1{U_{k+1}}+\frac1{U_k}\right)\nn\\%
&&\hspace{16mm}-\frac{2\al\beta\gm\de}{3\eta}\frac{3m+4}{\Ub U}%
+\frac{2\al\beta\gm\de}{\eta}\sum_{k=\phi}^{m}\frac{1}{U_{k+1}U_k},\label{qp3:s:leadeq}
\eeq
where $E_1$ is the integration constant of order of $1/\eta$. 

The explicit form of $E_1$ is derived by imposing the initial conditions on \eqref{qp3:s:leadeq}. Equation \eqref{qp3:s:leadeq} can be integrated by recognising that the coefficients of $\sbar_l$ and $s_l$ are exact differences, that is, we have
\beq
&&\hspace{-6mm}\frac{\Ub^2\left[U^2-(\gm+\de)U+\gm\de\right]-\gm\de\left[U^2-(\al+\beta)U+\al\beta\right]}{\Ub^2 U}=\frac{(\Ub-\Uu)(U-\gm)(U-\de)}{\Ub U},\nn\\
\label{qp3:sl:1}\\
\noalign{\textrm{and}}\nonumber\\
&&\hspace{-6mm}\frac{U^2\left[\Ub^2-(\gm+\de)\Ub+\gm\de\right]-\gm\de\left[\Ub^2-(\al+\beta)\Ub+\al\beta\right]}{\Ub U^2}=\frac{(U-\Ubb)(\Ub-\gm)(\Ub-\de)}{\Ub U},\nn\\
&&\vspace{-10mm}\label{qp3:sl:2}
\eeq
where we have used Equation \eqref{qp3:lead} in Equation \eqref{qp3:sl:1} and a forward iteration of \eqref{qp3:lead} is used in deducing \eqref{qp3:sl:2}.

Using the definition of $F$, we integrate Equation \eqref{qp3:s:leadeq} into the form \eqref{qp3:s:leadsol}, where the summation constant is zero. Note that for $s_l$ to be well-defined, we have assumed (w.l.o.g.) that $F$ is non-zero in the domain $\mathcal D_\delta$.
\end{proof}
{\remark $F$ can also be expressed as $\pm\sqrt{P(U,E_0}$, where
\beq
P(U,E_0)&=&(\gm -\de)^2U^4+2 \left(2\gm\de( \al+\beta+\gm+ \de)+(\gm+\de)  E_0\right)U^3\nn\\
&&+\left(E_0^2 -2\gm\de(\gm+\de)(\al+\beta)-4 \gm \de( \al\beta+\gm\de)\right)U^2+2 \gm  \de (2 \al \beta(\gm+\de)\nn\\
&& +(2\gm\de+E_0)(\al +\beta))U+\gm ^2 \de^2 (\al-\beta)^2.
\eeq}
{\cor\label{qp3:lem:Om} The approximate-period $\Om$ in the direction of $\om$ has the following expansion:
\be\label{qp3:Om}
\Om=\om-\frac{s(\phi+\om)}{U'(\phi)}
+\ord\left(\frac{1}{\eta^{2}}\right),
\ee
where $s=s_l$.}

\subsection{Evolution of the energy}
In this section, we calculate the local change of the energy across an approximate-period.

{\prop\label{qp3:prop:avg} The evolution of the energy $E$ is given by
\beq
&&E(\phi+\Om)-E(\phi)\label{qp3:avg}\nn\\
&=&\frac{(\al+\beta)\gm\de\2\om}{\eta}\left(\frac{1}{U(\phi+1)}+\frac{1}{U(\phi)}\right)%
-\frac{(\al+\beta)\gm\de}{\eta}\sum_{k=\phi+1}^{\phi+\Om}\left(\frac{1}{U_{k+1}}+\frac{1}{U_k}\right)\nn\\
&&\hspace{7mm}-\frac{2\al\beta\gm\de}{\eta}\frac{\om}{U(\phi+1)U(\phi)}%
+\frac{2\al\beta\gm\de}{\eta}\sum_{k=\phi+1}^{\phi+\Om}\frac{1}{U_{k+1} U_k}%
+\ord\left(\frac{1}{\eta^{2}}\right).
\eeq}
\begin{proof}
The proof can be shown by using Lemmas \ref{ubexpan:2nd} and \ref{ubexpan} and by following a similar argument as in the proof of Proposition \ref{qp1:avg}.
\end{proof}
We have now deduced all the results for the generic case of $q$-\pIII, which show that the energy is slowly varying of the order of $1/\eta$. We now turn to a special case of $q$-$\pIII$  to deduce explicit information about the elliptic function behaviour, in particular about the corresponding Picard-Fuchs equation. 

\subsection{Special case}\label{qp3:spe1}
In this section, we consider a special case of $q$-$\pIII$ with parameter values given by 
\begin{equation}\label{eq:sc}
\beta=-\al ,\ \de=-\gm,\ \al^2\neq\gm^2,\ \al, \gm\neq0. 
\end{equation}
{\lemma\label{qp3:U:sol1} In the special case given by Equation \eqref{eq:sc}, Equation (\ref{qp3:lead}) has the following explicit solution
\be
U(m)=\frac{k^{1/2}\sn(z)}{\mu},
\ee
where $z=\rho\2m$ and $\mu^2=\al\gm$, with periods given by
\be\label{qp3:om}
\om_i=\oint_{C_i}\frac{dU}{k^{1/2}\sqrt{\frac{1}{\al\gm}U^4-\frac{k^2+1}{k}U^2+\al\gm }},
\ee
where $i=1,\ 2$.}
\begin{proof}
This follows by explicit computation \cite{DLMF}.
\end{proof}
{\lemma Under the conditions \eqref{eq:sc}, the curve \eqref{qp3:E0} is parametrised by Jacobi $\sn$-functions.}
\begin{proof}
The first integral of Equation \eqref{qp3:lead} now takes the following form:
\be\label{qp3:inv:sp1}
\Ub^2U^2-E_0\Ub U-\gm^2\left(\Ub^2+U^2\right)+\al^2\gm^2=0.
\ee
The transformation $U=\mu V$, where $\mu$ is a constant, and $\mu^2=\pm \al\1\gm$ transforms \eqref{qp3:inv:sp1} into the canonical form of the biquadratic curve \cite{Baxt1982}.
By choosing $\mu^2=\al\,\!\gm$, we have
\beq
\bc
V&=\; k^{1/2}\sn(z)\\
\Vb&=k^{1/2}\sn(z+\rho_{\pm})
\ec,
\eeq
where
\beqn
z\;\;&=&\rho\2 m,\\
\sn^2\rho&=&\frac{\al}{\gm\,\!k_\pm},\\
k_{\pm}&=&\frac{-E_0^2+4\al^2\gm^2+4\gm^4 \pm \sqrt{\left(E_0^2-4a^2\gm^2-4\gm^4\right)^2-64a^2\gm^6 }}{8a\gm^3},
\eeqn
and $k_{+}=\frac{1}{k_{-}}$. Without loss of generality, we choose $k_{+}$.
\end{proof}
{\cor $V$ satisfies the following differential equation:
\beq
V_z^2&=&(kV^2-1)(V^2-k)=k\left(V^4+\frac{E_0^2-4\al^2\gm^2-4\gm^4}{4\al\gm^3}V^2+1\right).\label{qp3:Vdiff}
\eeq
Hence, $U$ satisfies the following differential equation:
\be\label{qp3:Udiff}
U_z^2=\frac{k}{\al\gm}U^4-\left(k^2+1\right)U^2+\al\gm k.
\ee}
The result follows by explicit computation.

{\lemma\label{qp3:sc1:PFE} The function $w=k^{1/2}\om$, where $\om$ is one of the periods $\om_i$, $i=1, 2$, satisfies the following differential equation as a function of $E$: 
\beq\label{qp3:sp1:wDE}
&&w_{EE}+\left(\frac{4E(E^2-4a^2\gm^2-4\gm^4)}{(E^2-4a^2\gm^2-4\gm^4)^2-64a^2\gm^6}-\frac1E\right)w_E\nn\\
&&\hspace{25mm}+\frac{E^2}{(E^2-4a^2\gm^2-4\gm^4)^2-64a^2\gm^6}w=0.
\eeq
\begin{proof}
See Appendix \ref{app:QP:PFE}.
\end{proof}
Equation \eqref{qp3:sp1:wDE} has five regular singular points, they are: $E=\pm 2\gm(\al+\gm)$, $\pm 2\gm(\al-\gm)$, and $\infty$ which are also the five degenerate points of \eqref{qp3:inv:sp1}. Under the identification $\zeta:=E^2$ the number of degenerate points reduces to three.}


{\remark We remark that by fixing $\al$, $\beta$, $\gm$, and let $\de$ go from $-\gm$ to $\gm$, the asymptotic limit of $U$ is a Weierstrass $\wp$-function. }

This concludes our proof for Theorem \ref{thm:qp3}.
\section{Conclusion}
In this paper, we derived the generic two parameter asymptotic behaviours of the solutions of two $q$-discrete Painlev\'e equations as the independent variable approaches infinity. For both equations it was shown that the asymptotic behaviours are determined by elliptic functions, in common with prior results on continuous Painlev\'e equations. We extended the method of averaging to these discrete equations and deduced the modulation of the corresponding energy-like parameter. For both equations, we showed that the energy $E$ varies slowly over the local period parallelogram. The combined results are stated as Theorems \ref{thm:qp1} and \ref{thm:qp3} in Section \ref{s:main}. 

In principle, the Picard-Fuchs equation, which governs the periods as functions of $E$, can be found for all values of the parameters. We have deduced an explicit Picard-Fuchs equation for a special case of parameter values in $q$-$\pIII$ here to provide an illutrative case. The result is used to estimate approximate-periods for the corresponding solutions. 

We note that the order of the slow variation of $E$ is analogous to the slow variation found for the continuous Painlev\'e equations \cite{JL2018}. This similarity suggests that there is a fundamental underlying connection  between these equations, possibly based on the geometry of their initial value spaces, which may be a fruitful direction for future research. More information on this correspondence may follow from applying our techniques to other discrete Painlev\'e equations.

\appendix
\section{Proof of Lemma \ref{ubexpan}}\label{ubexpan:pf}
In order to evaluate $u(\phi+\Om+1)$, we start by taking the Taylor expansion of $u(m)$ at $m=\phi+\om$, then we evaluate $m$ at $\phi+\Om$ and use the asymptotic expansion $u=U+s$:
\beqn
u(\phi+\Om)
&=&U(\phi)+s(\phi+\om)+(\Om-\om)\frac{d\, }{d\,\! n}\left(U(\phi)+s(\phi+\om)\right)\\
&&\phantom{U(\phi)+}+\frac{(\Om-\om)^2}{2}\frac{d^2\;}{d\,\!n^2}\left(U(\phi+\om)+s(\phi+\om)\right)+\ord\left((\Om-\om)^{-3}\right).
\eeqn
Recall that $u(\phi+\Om)=u(\phi)$ and by rearranging the above expansion, we obtain:
\be\label{dis:Om-om}
\Om-\om=\left.\left(-\frac{s}{U'}+\frac{ss'}{(U')^2}-\frac{s^2U'' }{2(U')^3}\right)\right|_{n=\phi+\om}+\ord(s^{3}),
\ee
where $'=\frac{d\;}{dn}$. This shows that $\Om-\om$ is of order $s$ as $|\eta|\to\infty$. This gives the $\Om$ expansion in the lemma. The next step is to Taylor expand $u(m)$ at $\phi+\om+1$ and evaluate at $\phi+\Om+1$. Then \eqref{dis:Om-om} is substituted into the resulting Taylor series:
\beqn
&&u(\phi+\Om+1)\\
&=&\Ub+\sbar-s\frac{\Ub'}{U'}+ss'\frac{\Ub'}{(U')^2}-s^2\frac{U'' \Ub'}{2(U')^3}-\frac{s\sbar'}{U'}\left.+s^2\frac{\Ub''}{(U')^2}\right|_{\phi+\om}+\ord\left(s^3\right).
\eeqn
The desired expansion for $u(\phi+\Om+1)$ follows after combining this result with that of the previous lemma. We note that in the proof we assumed that $s$, its iterations and its derivatives are all of the same order.

\section{Proof of Lemma \ref{qp3:sc1:PFE}}\label{app:QP:PFE}
Under the conditions \eqref{eq:sc}, we first change the variable of integration in the definition of $\om$ in Equation \eqref{qp3:om} into $V$, so that the period-integrals take the form 
\be\label{app:qp3:om}
\om_i=\frac{1}{k^{1/2}}\oint_{C_i}\frac{dV}{\sqrt{V^4+\frac{E^2-2\al^2\gm^2-4\gm^4}{4\al\gm^3}V^2+1}},
\ee
where $C_i$, $i=1,2$, are two independent contours on the Riemann surface. We define $w_i$ to be:
\be\label{app:qp3:w}
w_i:=\oint_{C_i}\frac{dV}{\sqrt{V^4+\frac{E^2-2\al^2\gm^2-4\gm^4}{4\al\gm^3}V^2+1}}.
\ee
For conciseness we derive the Picard-Fuchs equation for the independent variable $\zeta$, where $\zeta^{1/2}:=E$.
\beq
w=\oint\frac{dV}{\sqrt{V^4+\frac{\zeta-2\al^2\gm^2-4\gm^4}{4\al\gm^3}V^2+1}}.\label{app:qp3:w:eq}
\eeq
Using integration by parts on \eqref{app:qp3:w:eq}, we can express $w$ as:
\be\label{app:qp3:w:eq:2}
w=-2(\zeta-2\al^2\gm^2-4\gm^4)w_{\zeta}+2\Psi,
\ee
where
\beq
w_{\zeta}&=& -\frac{1}{8\al\gm^3}\oint\frac{V^2}{\left(V^4+\frac{\zeta-2\al^2\gm^2-4\gm^4}{4\al\gm^3}V^2+1\right)^{3/2}}dV,\text{ and}\\
\Psi&:=&\oint\frac{1}{\left(V^4+\frac{\zeta-2\al^2\gm^2-4\gm^4}{4\al\gm^3}V^2+1\right)^{3/2}}dV.
\eeq
Applying integration by parts on $\Psi$ gives the following relation:
\be\label{app:qp3:w:eq:3}
\Psi=128\2\al^2\gm^6\2w_{\zeta\zeta}-2\left(\zeta-2\al^2\gm^2-4\gm^4\right)\Psi_{\zeta},
\ee
where
\beq
w_{\zeta\zeta}&=& \frac{3}{64\al^2\gm^6}\oint\frac{V^4}{\left(V^4+\frac{\zeta-2\al^2\gm^2-4\gm^4}{4\al\gm^3}V^2+1\right)^{5/2}}dV,\text{ and}\\
\Psi_{\zeta}&=&-\frac{3}{8\al\gm^3}\oint\frac{V^2}{\left(V^4+\frac{\zeta-2\al^2\gm^2-4\gm^4}{4\al\gm^3}V^2+1\right)^{5/2}}dV.
\eeq
Combining \eqref{app:qp3:w:eq:2} and \eqref{app:qp3:w:eq:3} and eliminating any $\Psi$ dependence results in:
\be\label{app:qp3:PFE:w:zeta}
w_{\zeta\zeta}+\frac{2(\zeta-4\al^2\gm^2-4\gm^4)}{(\zeta-4\al^2\gm^2-4\gm^4)^2-64\al^2\gm^6}w_{\zeta}+\frac{w}{4\left[(\zeta-4\al^2\gm^2-4\gm^4)^2-64\al^2\gm^6\right]}=0,
\ee
which is equivalent to:
\beq
&&w_{EE}+\left(\frac{4E(E^2-4a^2\gm^2-4\gm^4)}{(E^2-4a^2\gm^2-4\gm^4)^2-64a^2\gm^6}-\frac1E\right)w_E\label{app:qp3:PFE:w:E}\\
&&\hspace{51mm}+\frac{E^2}{(E^2-4a^2\gm^2-4\gm^4)^2-64a^2\gm^6}w=0.\nn
\eeq
The Picard-Fuchs equation for $\om$ can be derived from this result by using the product rule.

\bibliographystyle{plain}
\bibliography{Thesis}

\end{document}